\documentclass[prl,twocolumn,preprintnumbers,amsmath,amssymb]{revtex4}

\usepackage{graphicx,color}
\usepackage{multirow}

\begin{document}


\title{Contact resistance dependence of crossed Andreev reflection}

\author{A. Kleine}
\email{andreas.kleine@unibas.ch}
\author{A. Baumgartner}
\author{J. Trbovic}
\author{C. Sch\"onenberger}
\affiliation{
Institute of Physics, University of Basel, Klingelbergstrasse 82, 4056 Basel, Switzerland\\}

\date{\today}

\begin{abstract}
We show experimentally that in nanometer scaled superconductor/normal metal hybrid devices and in a small window of contact resistances, crossed Andreev reflection (CAR) can dominate the nonlocal transport for all energies below the superconducting gap. Besides crossed Andreev reflection, elastic cotunneling (EC) and nonlocal charge imbalance can be identified as competing subgap transport mechanisms in temperature dependent four-terminal nonlocal measurements. We demonstrate a systematic change of the nonlocal resistance vs. bias characteristics with increasing contact resistances, which can be varied in the fabrication process. For samples with higher contact resistances, CAR is weakened relative to EC in the midgap regime, possibly due to dynamical Coulomb blockade. Gaining control of crossed Andreev reflection is an important step towards the realization of a solid state entangler.

\end{abstract}


\maketitle

Quantum mechanically entangled pairs of particles are a major building block of quantum computation and information processing. A natural source of entangled electrons is a BCS-type superconductor where the Cooper pairs (CP) form spin singlet states. The two electrons of a Cooper pair can be spatially separated into two different metallic leads in a nonlocal process called crossed Andreev reflection (CAR) \cite{Recher_Loss_PRB63_2001, Lesovik_EPJB24_2001, Byers_PRL74_1995, Deutscher_APL76_2000}.

\begin{figure}[b]{
\centering
\includegraphics{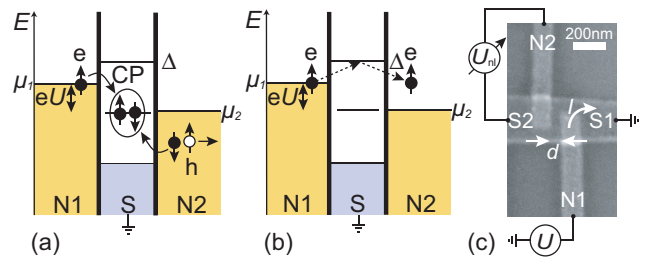}
}
\caption{(Color online) Schematics of (a) crossed Andreev reflection and (b) elastic cotunneling. (c) SEM image of a typical sample. A current $I$ is applied between normal contact N1 and the superconducting contact S1, while the nonlocal voltage $U_{\text{nl}}$ is detected between N2 and S2.
\label{Figure1}}
\end{figure}

At temperatures ($T$) well below the superconducting transition temperature, $T_{\text{c}}$, and for bias potentials below the superconducting energy gap $\Delta$, electrons from a normal metal contact (N) can enter the superconductor (S) only as Cooper pairs by a process known as Andreev reflection (AR). In this local process a hole is reflected into the same N to conserve momentum. If two normal metal contacts, N1 and N2, are spatially separated by less than the coherence length $\xi$, the two electrons forming a CP can originate from different normal contacts, see Fig.~\ref{Figure1}(a). This process opens an additional nonlocal conduction path known as CAR. An inverse process was proposed as the basis of a solid-state entangler: the electrons of a CP are split between the two leads while retaining their entanglement from the superconductor. However, this method of creating entangled particles can be accompanied by two additional processes that lead to correlated signals on N1 and N2, but not to entanglement. In the first, a single electron from N1 can reach the other contact N2 by elastic cotunneling (EC)  \cite{Averin_PRL65_1990, Falci_EPL54_2001, Melin_PRB70_2004}, see Fig.~\ref{Figure1}(b). In the second, called nonlocal charge imbalance (CI), electrical charge can be transferred to the second contact by the diffusion of quasi-particles generated by finite temperatures or finite bias.

Recently, the relative strength of these subgap processes was the subject of extensive theoretical work. Standard BCS theory predicts that to lowest order in the tunneling rates, CAR and EC exactly cancel in normal metal/insulator/superconductor (NIS) systems at low $T$ and bias \cite{Falci_EPL54_2001}. This cancelation can be lifted for higher transmissions \cite{Kalenkov_PRB75_2007}, by spin-active interfaces \cite{Kalenkov_PRB76_2007} or ferromagnetic contacts \cite{Melin_PRB70_2004}, by disorder, or by electron-electron interactions \cite{Morten_PRB74_2006, Duhot_PRB75_2007, Golubev_PRB76_2007}. It has been suggested that Coulomb interaction and the electromagnetic environment are crucial for CAR to be the dominant process \cite{Yeyati_naturephysics3_2007}.

Experimentally, first signatures of nonlocal transport processes have been reported only recently on multi-terminal hybrid structures. We can group the experiments into two categories: while EC and CI dominate the subgap transport in devices with highly transparent contacts \cite{Beckmann_PRL93_2004,Cadden-Zimansky_Chandrasekhar_PRL97_2006}, CAR is more prominent in samples with lower transparencies, depending also on the applied bias \cite{Russo_Klapwijk_PRL95_2005, Beckmann_ApplPhysA89_2007}. These individual findings suggest that the contact resistance plays a significant role in determining which process is dominant.

Here we present four-terminal nonlocal experiments on a series of samples of similar geometry but with NIS junctions of different resistances and with non-magnetic contacts to avoid stray field effects. We demonstrate that in a small window of contact resistances CAR can dominate the nonlocal transport for all subgap bias voltages at low temperatures, and that CAR is suppressed for larger barrier resistances.

In our experiments we perform four-terminal nonlocal measurements: a bias $U$ between the normal metal contact N1 (injector) and the  grounded superconductor lead S1 causes a current $I$ as shown in Fig.~\ref{Figure1}(c). We measure both, $I$ and the nonlocal potential difference $U_{\text{nl}}$ between the normal contact N2 (detector) and the second superconducting contact S2. In a CAR process a particle is injected from N1 and one of opposite charge is generated in N2. In Fig.~\ref{Figure1}(a) such a process is shown schematically for $U<0$, where electrons are injected from N1 and holes are created in N2. For each $U$ a potential of {\it opposite} sign builds up on N2 to ensure the zero current boundary condition of a voltage measurement. In EC and CI processes, however, the charges on the injector and detector have the same sign and $U_{\text{nl}}$ has the same polarity as the bias. In practice, we use ac modulation techniques and investigate the nonlocal differential resistance $R_{\text{nl}}=dU_{\text{nl}}/dI\approx U^{\text{ac}}_{\text{d}}/I^{\text{ac}}_{\text{i}}$, where 'ac' denotes the amplitude of the corresponding variable. $R_{\text{nl}}$ is negative for CAR and positive for all other subgap processes.

The samples are planar multi-terminal hybrid structures, see Fig.~\ref{Figure1}(c). They were fabricated on thermally oxidized Si wafers by e-beam lithography and using the angle-evaporation technique in ultra-high vacuum with a base pressure of $10^{-10}\,$mbar. Because of the low spin-orbit interaction, we chose Al as superconducting material. The Al layer has a thickness of $50\,$nm and a width of $200$-$300\,$nm. We oxidized the Al between $3$ and $15\,$min {\it in situ} in an oxygen atmosphere with a pressure in the range of $0.1$ to $12\,$mbar to obtain the required tunnel barriers. Without breaking the vacuum, we tilted the sample and deposited Pd as normal metal contacts with a typical thickness of $30$-$40\,$nm and a width of $90\,$nm. In our Al/AlO$_x$/Pd-samples we obtained edge-to-edge distances $d$ between the N electrodes of $\sim50-220\,$nm.

For each device the characteristic energy $\Delta^{*}$ denotes the bias potential of maximum conductance of the injector contact in superconducting tunneling spectroscopy (STS) measurements, see Table I. We note that $\Delta^{*}$ depends on the barrier characteristics and does not coincide with the bulk energy gap of $\Delta\approx 200\,\mu$eV. We measured a critical temperature $T_{\text{c}}\approx1.2\,$K and a resistivity $\rho_{\text{Al}}=1.2-2.6\,\mu\Omega\text{cm}$ of the Al layers at $T=4.2\,$K. We find an elastic mean free path $\ell_{\text{el}}\approx 20-50\,$nm and $\xi\approx 180-270\,$nm at cryogenic temperatures \footnote{We use $\xi\cong\sqrt{\xi_{0}\ell_{\text{el}}}$, $\xi_0=\hbar v_F/\pi\Delta(0)$, $\ell_{\text{el}}=3D/v_F$ and the diffusion constant $D=\sigma_{\text{Al}}/e^{2}N_{\text{Al}}$. $N_{\text{Al}}=2.4\cdot10^{28}\,$$1/\text{eVm}^3$ is the density of states of Al at the Fermi energy, $v_F=1.3\cdot10^6\,$m/s the Fermi velocity and $\sigma_{\text{Al}}$ the conductivity of the Al layer.}. Therefore the Al is in the 'dirty limit' of superconductivity.

\begin{figure}[b]{
\centering
\includegraphics{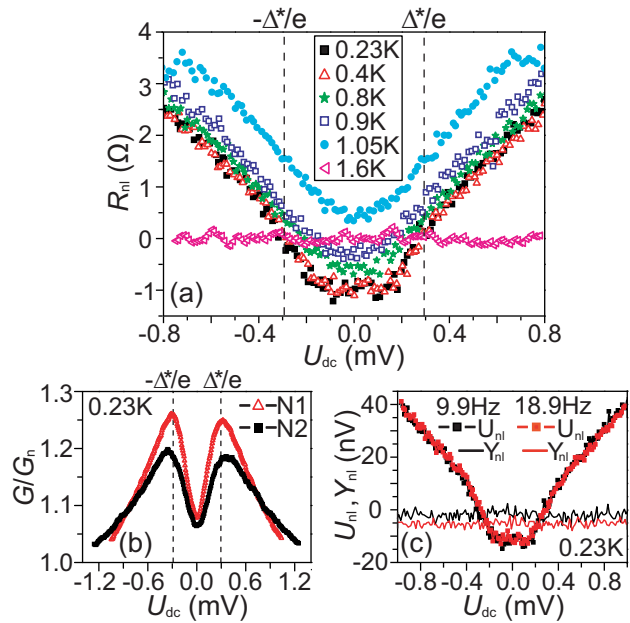}
}
\caption{(Color online) Results of sample A. (a) $R_{\text{nl}}$ vs. bias for various temperatures with N1 as injector and N2 as detector. (b) STS characterization of contacts N1 and N2 at $0.23\,$K. The dashed lines label the bias corresponding to $\Delta^{*}$ of the injector. (c) Frequency dependence of the in-phase nonlocal signal $U_{\text{nl}}$ and the capacitive part $Y_{\text{nl}}$.
\label{Figure2}}
\end{figure}

We performed nonlocal ac measurements using lock-in amplifiers at a frequency of $\sim10\,$Hz in a He$^{3}$-cryostat with a base temperature of $0.23\,$K. The dc bias applied to N1, $U_{\text{dc}}$, is modulated by $U^{\text{ac}}_{\text{i}}\approx 12\,\mu$Vrms. Since the junction resistances vary strongly with the applied bias, leakage currents and capacitive cross-talk can produce spurious signals which might resemble the expected characteristics. To exclude cross-talk we repeated all measurements at various frequencies. For all samples shown here we find that the in-phase part of $U_{\text{nl}}$ is independent of the measurement frequency, while the capacitive part $Y_{\text{nl}}$ is negligible, see Fig.~\ref{Figure2}(c). For comparison we simultaneously measure $U_{\text{nl}}$ in a DC configuration with $10\,$s delay between points (not shown). The dc coincide with the ac results, but suffer from a much worse signal-to-noise ratio. Another source of spurious $U_{\text{nl}}$ signals is resistive leakage into the detector. We reduce this leakage by using voltage amplifiers with 1~G$\Omega$ input impedance and check for each device that a change to $100\,$M$\Omega$ has no effect on our results. We note that for $T>T_{\text{c}}$ all nonlocal signals vanish, implying that they are related to the superconductor and not to the measurement setup or inhomogeneous current paths.

\begin{table}
	\caption{Properties of samples A, B and C. $\ell_{\text{el}}$: elastic mean free path of the Al strip at $T=4.2\,$K; $\xi$: superconducting coherence length; $R_{\text{i}}$, $R_{\text{d}}$: injector and detector barrier resistance at $T=1.5\,$K; $\Delta^{*}$: characteristic energy of the injector; d: injector-detector distance; $R_{\text{i}}A_{\text{i}}$, $R_{\text{d}}A_{\text{d}}$: resistance area product of the injector and detector; asymmetry  $\alpha=(R_{\text{d}}A_{\text{d}}-R_{\text{i}}A_{\text{i}})/(R_{\text{d}}A_{\text{d}}+R_{\text{i}}A_{\text{i}})$.}
			\label{table1}
			\begin{tabular}{cccccccccc}\hline\hline
 			
   & $\ell_{\text{el}}$  &  $\xi$ & $R_{\text{i}}$ & $R_{\text{d}}$ & $\Delta^{*}$ & $d $ & $R_{\text{i}}A_{\text{i}}$ & $R_{\text{d}}A_{\text{d}}$ & $\alpha$ \\
	  & [nm] & [nm]  & [k$\Omega$]& [k$\Omega$] & [meV] & [nm] & [$\Omega\mu\text{m}^2$] & [$\Omega\mu\text{m}^2$] & \\\hline												 
A   & 23  & 184 &  0.4 & 0.5  & 0.29 & 55 & 8.5 & 16.9 & 0.33 \\

B   & 24  & 189  & 0.8 & 5.5  & 0.26 & 155 & 22.1 & 93.5 & 0.62 \\

C   & 50  & 273 & 2.6  & 11.2  & 0.32 & 220 & 116.6 & 365.4 & 0.52

 \\\hline\hline

	\end{tabular}
\end{table}
In the following we present experiments on three samples A, B, and C, which have a similar geometry, but differ in the contact resistances. The characteristics of each sample are summarized in Table~\ref{table1}.

Figure~\ref{Figure2}(a) shows $R_{\text{nl}}$ as a function of the bias voltage $U_{\text{dc}}$ at various temperatures for sample A. At base temperature we observe a negative nonlocal resistance $R_{\text{nl}}$ for all subgap voltages with $R_{\text{nl}}\approx-1.0\,\Omega$. This value does not change with temperature up to $T\approx0.5\,$K. When increasing $T$ further $R_{\text{nl}}$ increases and becomes positive for $T\geq1.05\,$K. For voltages larger than $\Delta^{*}/e$, $R_{\text{nl}}$ is positive at all temperatures and increases with $T$. At $1.6\,$K, i.e. for $T>T_{\text{c}}$, $R_{\text{nl}}\approx0$ for all biases.  The contact characterization of sample A at base temperature is shown in Fig.~\ref{Figure2}(b) and exhibits a significant subgap conductance for both, N1 and N2. The dashed lines represent the bias corresponding to the energy $\Delta^{*}$ of the injector, which coincides with the sign change of $R_{\text{nl}}$. Within the energy gap both barriers exhibit nearly the same normalized conductance values and both contacts have similar resistances above $T_{\text{c}}$, $0.4\,$k$\Omega$ and $0.5\,$k$\Omega$. Figure~\ref{Figure2}(c) shows the frequency dependence of the nonlocal signal as discussed above.

\begin{figure}[b]{
\centering
\includegraphics{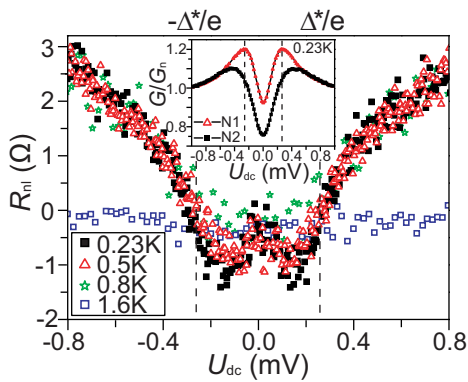}
}
\caption{(Color online) Results of sample B. $R_{\text{nl}}$ vs. bias at various $T$ with N1 as injector and N2 as detector. The inset shows the STS characterization of N1 and N2 at $0.23\,$K. The dashed lines indicate the bias corresponding to $\Delta^{*}$ of the injector.
\label{Figure3}}
\end{figure}

In Fig.~\ref{Figure3} the bias dependence of $R_{\text{nl}}$ of sample B is plotted for several temperatures. The shape and amplitudes of the curves are very similar to those of sample A in Fig.~\ref{Figure2}(a), as is the temperature dependence. However, in sample B at base temperature we observe a local maximum for $U_{\text{dc}}\approx0$ with $R_{\text{nl}}\approx0$, whereas for finite subgap biases we find $R_{\text{nl}}<0$ with two minima of $R_{\text{nl}}\approx-1.1\,\Omega$ at $U_{\text{dc}}\approx\pm0.235\,$mV. Above $T_{\text{c}}$ the nonlocal signals vanish for all bias voltages. The inset shows the STS characterization of the individual contacts at $0.23\,$K. In contrast to sample A, the normalized injector and detector conductances differ considerably in the gap and are smaller than in sample A ($R_{\text{i}}=0.8\,$k$\Omega$ and $R_{\text{d}}=5.5\,$k$\Omega$).

\begin{figure}[b]{
\centering
\includegraphics{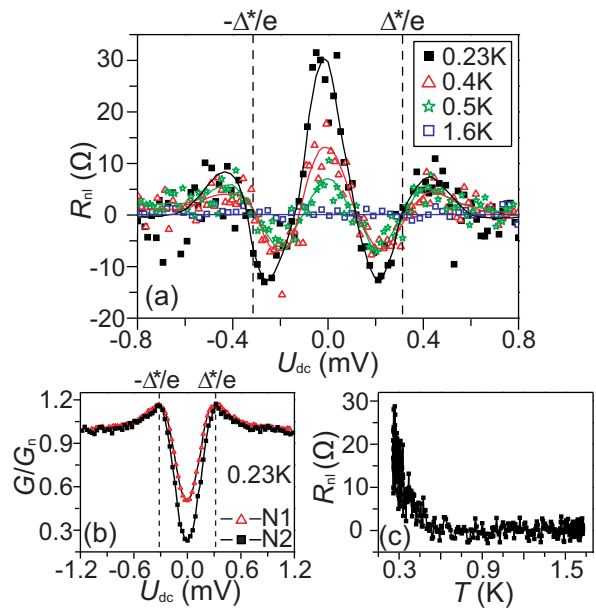}
}
\caption{(Color online) Results of sample C. (a) $R_{\text{nl}}$ vs. bias for various $T$ with N1 as injector and N2 as detector. The solid lines are guides to the eye and the dashed lines show the voltage corresponding to $\Delta^{*}$ of the injector. (b) STS-measurements of contact N1 and N2 at $0.23\,$K. (c) $R_{\text{nl}}$ at zero bias as a function of $T$.
\label{Figure4}}
\end{figure}

Bias dependent measurements for sample C are shown in Fig.~\ref{Figure4}(a). At base temperature a prominent local maximum in the middle of the superconducting gap develops with a {\it positive} maximum of $R_{\text{nl}}\approx30\,\Omega$. For $0.13\,\text{mV}\alt\left|U_{\text{dc}}\right|\alt\Delta^{*}/e$ with $\Delta^{*}\approx0.31\,$meV, $R_{\text{nl}}$ is negative with minima at $\pm0.23\,$mV with $R_{\text{nl}}\approx-13\,\Omega$. The sign changes at $U_{\text{dc}}\approx\pm0.13\,$mV, independently of temperature. For $\left|U_{\text{dc}}\right|>\Delta^{*}/e$ the signal is positive and tends to zero for increasing bias, in contrast to samples A and B.
With increasing temperature all nonlocal signals tend to zero, independently of the bias, with $R_{\text{nl}}\approx0$ for $T>T_{\text{c}}$. At zero bias the signal decreases monotonically with increasing $T$ and vanishes already for $T>0.5\,$K, see Fig.~\ref{Figure4}(c). A finite nonlocal signal only develops for $T$ well below $T_{\text{c}}$. From the STS measurements shown in Fig.~\ref{Figure4}(b) we find that the two contact resistances are quite different and much larger than in the previous samples ($R_{\text{i}}=2.6\,$k$\Omega$ and $R_{\text{d}}=11.2\,$k$\Omega$).

Based on the discussion above we interpret $R_{\text{nl}}<0$ in the superconducting energy gap as CAR being the dominant process, while $R_{\text{nl}}>0$ is attributed to EC or CI. The latter two processes can be distinguished by the temperature dependence of $R_{\text{nl}}$ at a given bias. EC and CAR are reduced with increasing temperature \cite{Yeyati_naturephysics3_2007, Russo_Klapwijk_PRL95_2005}, while CI increases strongly up to $T_{\text{c}}$ due to the decreased superconducting energy gap and the broadened thermal distribution, and vanishes for $T>T_{\text{c}}$ \cite{Cadden-Zimansky_Chandrasekhar_PRL97_2006}. At zero bias and low enough temperatures CI can be neglected.

For sample A we conclude that CAR is the dominant nonlocal transport process at base temperature and at all subgap biases. From the characteristic strong increase of $R_{\text{nl}}$ to positive values with $T$, we conclude that nonlocal CI becomes important for $T>0.5\,$K. These findings also apply for sample B, except around zero bias, where CAR and EC approximately compensate each other at low temperatures. At finite subgap bias, EC is weakened as reported in \cite{Russo_Klapwijk_PRL95_2005} and the characteristics become similar to sample A. In sample C, EC not only compensates CAR, but even dominates the nonlocal midgap transport. As in A and B, CAR dominates at finite subgap bias. In all samples CI dominates for bias potentials above the energy gap. In contrast to the first two samples, CI is strongly suppressed in sample C for $\left|U_{\text{dc}}\right|>>\Delta^{*}/e$.

Our experimental results on different samples allow us to examine which device parameter determines the $R_{\text{nl}}$ characteristics. We now discuss the following possible parameters: 1) the distance $d$ between injector and detector 2) the resistance of the injector and detector contacts 3) the resistance-area product, RA, of the contacts, and 4) the RA asymmetry $\alpha$ between injector and detector. First we note that $d$ does not change the characteristic shape of the $R_{\text{nl}}$ curve \cite{Russo_Klapwijk_PRL95_2005}. In addition, it is well established, both theoretically and experimentally, that all nonlocal subgap signals at low bias are reduced monotonically with $d$ \cite{Falci_EPL54_2001,Russo_Klapwijk_PRL95_2005,Beckmann_PRL93_2004,Cadden-Zimansky_Chandrasekhar_PRL97_2006}. Since the junction separation is largest in sample C, which exhibits the largest nonlocal signals, we conclude that as long as $d<\xi$, the characteristics of the nonlocal response is determined by another parameter. Our results suggest that larger resistances of the injector and detector contacts suppress CAR and favor EC. A comparison with similar experiments in the literature \cite{Russo_Klapwijk_PRL95_2005,Beckmann_ApplPhysA89_2007} yields $R_{\text{nl}}$ curves consistent with our data when we use RA as the relevant parameter. This is remarkable since in \cite{Russo_Klapwijk_PRL95_2005} the geometry and materials are very different. In contrast to RA, we find no correlation between the RA asymmetry $\alpha=(R_{\text{d}}A_{\text{d}}-R_{\text{i}}A_{\text{i}})/(R_{\text{d}}A_{\text{d}}+R_{\text{i}}A_{\text{i}})$ and the systematic evolution of our data from sample A to sample C. From this discussion we conclude that the shape of the $R_{\text{nl}}$ curves of our samples is determined by RA.

Our qualitative explanation for this dependence is based on a model in which the two normal contacts are electromagnetically coupled, which can favor CAR over EC and lead to $R_{\text{nl}}<0$, depending on the symmetry of the coupling \cite{Yeyati_naturephysics3_2007}. This model requires dynamical Coulomb blockade (DCB) on the two NS junctions in union with the complex impedance of the S contacts. This leads to a blockade on the superconductor which suppresses CAR, because the net charge $-2e$ of a CP is added to the superconductor and the corresponding charging energy has to be supplied. Since CAR also vanishes for completely transparent contacts \cite{Kalenkov_PRB75_2007}, this leaves only a small window of transparencies where CAR can be observed at zero bias. In contrast to CAR, no net charge is added for EC, which is therefore essentially unaffected by the blockade and dominates the subgap transport for low transparencies around zero bias. 

DCB depends strongly on the environmental impedance and thus on the resistance of the tunneling contacts, $R_{\text{T}}$ \cite{Devoret_PRL64_1990}. It was shown that DCB can occur also for $R_{\text{T}}<h/2e^2\approx12\,\text{k}\Omega$, and that decreasing $R_{\text{T}}$ leads to a reduction of the DCB effects \cite{Joyez_PRL80_1998}. RA is the natural parameter to characterize a finite area tunnel junction with many parallel channels since it scales with the average channel resistance. A large number of weakly transmitting channels may lead to a small $R_{\text{T}}$ while retaining DCB for each channel \cite{Joyez_PRL80_1998}.

The systematic change of the measured nonlocal resistance vs. bias characteristics in our samples can thus be understood as follows: the contacts of sample A have the lowest resistances and the blockade is lifted, which allows CI and CAR to develop. The latter is favored over EC possibly due to an electromagnetic coupling of the injector and detector characteristic for all our samples. For sample B with slightly larger contact resistance, DCB sets in and weakens the dominance of CAR around zero bias which leads to a reduction of the negative nonlocal resistance. At a finite subgap bias EC is weaker because the process can not be elastic anymore. With resistances of a few k$\Omega$ as in sample C, DCB becomes strong and inhibits CAR relative to EC. We note that (local) AR is suppressed stronger in the DCB regime than CAR on two independent (but spatially close) contacts \cite{Recher_PRL91_2003}, which might explain the larger $R_{\text{nl}}$ values in sample C. Since for CI a net charge is transferred to the superconductor, the DCB picture is also consistent with the suppression of CI in sample C at large bias.

In conclusion we find in our experiments that for small RA CAR can dominate the transport characteristics for all subgap biases, whereas above the superconducting energy gap the CI rates are important. We have shown that CAR is suppressed for larger RA around zero bias. We qualitatively explain the various subgap transport characteristics by different dependences of the relevant processes on dynamical Coulomb blockade, which depends crucially on the RA.

We thank A. D. Zaikin, D. S. Golubev and A. Levy Yeyati for fruitful discussions. This work is financially supported by the NCCR on Nanoscale Science and the Swiss-NSF.

\bibliographystyle{apsrev}

\begin{thebibliography}{99}
\expandafter\ifx\csname natexlab\endcsname\relax\def\natexlab#1{#1}\fi
\expandafter\ifx\csname bibnamefont\endcsname\relax
  \def\bibnamefont#1{#1}\fi
\expandafter\ifx\csname bibfnamefont\endcsname\relax
  \def\bibfnamefont#1{#1}\fi
\expandafter\ifx\csname citenamefont\endcsname\relax
  \def\citenamefont#1{#1}\fi
\expandafter\ifx\csname url\endcsname\relax
  \def\url#1{\texttt{#1}}\fi
\expandafter\ifx\csname urlprefix\endcsname\relax\def\urlprefix{URL }\fi
\providecommand{\bibinfo}[2]{#2}
\providecommand{\eprint}[2][]{\url{#2}}

\bibitem{Byers_PRL74_1995}
\bibinfo{author}{\bibfnamefont{J.M.}~\bibnamefont{Byers}},
\bibnamefont{and}
\bibinfo{author}{\bibfnamefont{M.E.}~\bibnamefont{Flatt\'{e}}},
\bibinfo{journal}{Phys. Rev. Lett.}
\textbf{\bibinfo{volume}{74}},
\bibinfo{pages}{306}
(\bibinfo{year}{1995}).

\bibitem{Deutscher_APL76_2000}
\bibinfo{author}{\bibfnamefont{G.}~\bibnamefont{Deutscher}},
\bibnamefont{and}
\bibinfo{author}{\bibfnamefont{D.}~\bibnamefont{Feinberg}},
\bibinfo{journal}{Appl. Phys. Lett.}
\textbf{\bibinfo{volume}{76}},
\bibinfo{pages}{487}
(\bibinfo{year}{2000}).

\bibitem{Lesovik_EPJB24_2001}
\bibinfo{author}{\bibfnamefont{G.B.}~\bibnamefont{Lesovik}},
\bibinfo{author}{\bibfnamefont{T.}~\bibnamefont{Martin}},
\bibnamefont{and}
\bibinfo{author}{\bibfnamefont{G.}~\bibnamefont{Blatter}},
\bibinfo{journal}{Eur. Phys. J. B}
\textbf{\bibinfo{volume}{24}},
\bibinfo{pages}{287}
(\bibinfo{year}{2001}).

\bibitem{Recher_Loss_PRB63_2001}
\bibinfo{author}{\bibfnamefont{P.}~\bibnamefont{Recher}},
\bibinfo{author}{\bibfnamefont{E.V.}~\bibnamefont{Sukhorukov}},
\bibnamefont{and}
\bibinfo{author}{\bibfnamefont{D.}~\bibnamefont{Loss}},
\bibinfo{journal}{Phys. Rev. B}
\textbf{\bibinfo{volume}{63}},
\bibinfo{pages}{165314}
(\bibinfo{year}{2001}).


\bibitem{Averin_PRL65_1990}
\bibinfo{author}{\bibfnamefont{D.V.}~\bibnamefont{Averin}},
\bibnamefont{and}
\bibinfo{author}{\bibfnamefont{Y.V.}~\bibnamefont{Nazarov}},
\bibinfo{journal}{Phys. Rev. Lett.}
\textbf{\bibinfo{volume}{65}},
\bibinfo{pages}{2446}
(\bibinfo{year}{1990}).

\bibitem{Falci_EPL54_2001}
\bibinfo{author}{\bibfnamefont{G.}~\bibnamefont{Falci}},
\bibinfo{author}{\bibfnamefont{D.}~\bibnamefont{Feinberg}},
\bibnamefont{and}
\bibinfo{author}{\bibfnamefont{F.W.J.}~\bibnamefont{Hekking}},
\bibinfo{journal}{Europhys. Lett.}
\textbf{\bibinfo{volume}{54}},
\bibinfo{pages}{255}
(\bibinfo{year}{2001}).

\bibitem{Melin_PRB70_2004}
\bibinfo{author}{\bibfnamefont{R.}~\bibnamefont{M\'{e}lin}},
\bibnamefont{and}
\bibinfo{author}{\bibfnamefont{D.}~\bibnamefont{Feinberg}},
\bibinfo{journal}{Phys. Rev. B}
\textbf{\bibinfo{volume}{70}},
\bibinfo{pages}{174509}
(\bibinfo{year}{2004}).

\bibitem{Kalenkov_PRB75_2007}
\bibinfo{author}{\bibfnamefont{M.S.}~\bibnamefont{Kalenkov}},
\bibnamefont{and}
\bibinfo{author}{\bibfnamefont{A.D.}~\bibnamefont{Zaikin}},
\bibinfo{journal}{Phys. Rev. B}
\textbf{\bibinfo{volume}{75}},
\bibinfo{pages}{172503}
(\bibinfo{year}{2007}).

\bibitem{Kalenkov_PRB76_2007}
\bibinfo{author}{\bibfnamefont{M.S.}~\bibnamefont{Kalenkov}},
\bibnamefont{and}
\bibinfo{author}{\bibfnamefont{A.D.}~\bibnamefont{Zaikin}},
\bibinfo{journal}{Phys. Rev. B}
\textbf{\bibinfo{volume}{76}},
\bibinfo{pages}{224506}
(\bibinfo{year}{2007}).

\bibitem{Morten_PRB74_2006}
\bibinfo{author}{\bibfnamefont{J.P.}~\bibnamefont{Morten}},
\bibinfo{author}{\bibfnamefont{A.}~\bibnamefont{Brataas}},
\bibnamefont{and}
\bibinfo{author}{\bibfnamefont{W.}~\bibnamefont{Belzig}},
\bibinfo{journal}{Phys. Rev. B}
\textbf{\bibinfo{volume}{74}},
\bibinfo{pages}{214510}
(\bibinfo{year}{2006}).


\bibitem{Duhot_PRB75_2007}
\bibinfo{author}{\bibfnamefont{S.}~\bibnamefont{Duhot}},
\bibnamefont{and}
\bibinfo{author}{\bibfnamefont{R.}~\bibnamefont{M\'{e}lin}},
\bibinfo{journal}{Phys. Rev. B}
\textbf{\bibinfo{volume}{75}},
\bibinfo{pages}{184531}
(\bibinfo{year}{2007}).


\bibitem{Golubev_PRB76_2007}
\bibinfo{author}{\bibfnamefont{D.S.}~\bibnamefont{Golubev}},
\bibnamefont{and}
\bibinfo{author}{\bibfnamefont{A.D.}~\bibnamefont{Zaikin}},
\bibinfo{journal}{Phys. Rev. B}
\textbf{\bibinfo{volume}{76}},
\bibinfo{pages}{184510}
(\bibinfo{year}{2007}).


\bibitem{Yeyati_naturephysics3_2007}
\bibinfo{author}{\bibfnamefont{A.}~\bibnamefont{Levy Yeyati}},
\bibinfo{author}{\bibfnamefont{F.S.}~\bibnamefont{Bergeret}},
\bibinfo{author}{\bibfnamefont{A.}~\bibnamefont{Martin-Rodero}},
\bibnamefont{and}
\bibinfo{author}{\bibfnamefont{T.M.}~\bibnamefont{Klapwijk}},
\bibinfo{journal}{Nature Phys.}
\textbf{\bibinfo{volume}{3}},
\bibinfo{pages}{455}
(\bibinfo{year}{2007}).

\bibitem{Beckmann_PRL93_2004}
\bibinfo{author}{\bibfnamefont{D.}~\bibnamefont{Beckmann}},
\bibinfo{author}{\bibfnamefont{H.B.}~\bibnamefont{Weber}},
\bibnamefont{and}
\bibinfo{author}{\bibfnamefont{H.}~\bibnamefont{v. L\"ohneysen}},
\bibinfo{journal}{Phys. Rev. Lett.}
\textbf{\bibinfo{volume}{93}},
\bibinfo{pages}{197003}
(\bibinfo{year}{2004}).

\bibitem{Cadden-Zimansky_Chandrasekhar_PRL97_2006}
\bibinfo{author}{\bibfnamefont{P.}~\bibnamefont{Cadden-Zimansky}},
\bibnamefont{and}
\bibinfo{author}{\bibfnamefont{V.}
~\bibnamefont{Chandrasekhar}},
\bibinfo{journal}{Phys. Rev. Lett.}
\textbf{\bibinfo{volume}{97}},
\bibinfo{pages}{237003}
(\bibinfo{year}{2006}).

\bibitem{Russo_Klapwijk_PRL95_2005}
\bibinfo{author}{\bibfnamefont{S.}~\bibnamefont{Russo}},
\bibinfo{author}{\bibfnamefont{M.}~\bibnamefont{Kroug}},
\bibinfo{author}{\bibfnamefont{T.M.}~\bibnamefont{Klapwijk}},
\bibnamefont{and}
\bibinfo{author}{\bibfnamefont{A.F.}~\bibnamefont{Morpurgo}},
\bibinfo{journal}{Phys. Rev. Lett.}
\textbf{\bibinfo{volume}{95}},
\bibinfo{pages}{027002}
(\bibinfo{year}{2005}).

\bibitem{Beckmann_ApplPhysA89_2007}
\bibinfo{author}{\bibfnamefont{D.}~\bibnamefont{Beckmann}},
\bibnamefont{and}
\bibinfo{author}{\bibfnamefont{H.}~\bibnamefont{v. L\"ohneysen}},
\bibinfo{journal}{Appl. Phys. A}
\textbf{\bibinfo{volume}{89}},
\bibinfo{pages}{603}
(\bibinfo{year}{2007}).

\bibitem{Devoret_PRL64_1990}
\bibinfo{author}{\bibfnamefont{M.H.}~\bibnamefont{Devoret}},
\bibinfo{author}{\bibfnamefont{D.}~\bibnamefont{Esteve}},
\bibinfo{author}{\bibfnamefont{H.}~\bibnamefont{Grabert}},
\bibinfo{author}{\bibfnamefont{G.-L.}~\bibnamefont{Ingold}},
\bibinfo{author}{\bibfnamefont{H.}~\bibnamefont{Pothier}},
\bibnamefont{and}
\bibinfo{author}{\bibfnamefont{C.}~\bibnamefont{Urbina}},
\bibinfo{journal}{Phys. Rev. Lett.}
\textbf{\bibinfo{volume}{64}},
\bibinfo{pages}{1824}
(\bibinfo{year}{1990}).

\bibitem{Joyez_PRL80_1998}
\bibinfo{author}{\bibfnamefont{P.}~\bibnamefont{Joyez}},
\bibinfo{author}{\bibfnamefont{D.}~\bibnamefont{Esteve}},
\bibinfo{author}{\bibfnamefont{M.H.}~\bibnamefont{Devoret}},
\bibinfo{journal}{Phys. Rev. Lett.}
\textbf{\bibinfo{volume}{80}},
\bibinfo{pages}{1956}
(\bibinfo{year}{1998}).


\bibitem{Recher_PRL91_2003}
\bibinfo{author}{\bibfnamefont{P.}~\bibnamefont{Recher}},
\bibnamefont{and}
\bibinfo{author}{\bibfnamefont{D.}~\bibnamefont{Loss}},
\bibinfo{journal}{Phys. Rev. Lett.}
\textbf{\bibinfo{volume}{91}},
\bibinfo{pages}{267003}
(\bibinfo{year}{2003}).

\end{thebibliography}

\end{document}